\documentclass[amsmath,amssymb,floatfix,lengthcheck,nopreprintnumbers,prl,showpacs,twocolumn]{revtex4}
\usepackage{graphicx}

\newcommand{\beq}{\begin{equation}}
\newcommand{\eeq}{\end{equation}}
\newcommand{\beqs}{\begin{eqnarray}}
\newcommand{\eeqs}{\end{eqnarray}}

\begin{document}

\title{Lattice Study of the Conformal Window in QCD-like Theories}

\author{Thomas Appelquist}
\affiliation{Department of Physics, Sloane Laboratory, Yale University,
             New Haven, CT, 06520}

\author{George T.\ Fleming}
\affiliation{Department of Physics, Sloane Laboratory, Yale University,
             New Haven, CT, 06520}

\author{Ethan T.\ Neil}
\affiliation{Department of Physics, Sloane Laboratory, Yale University,
             New Haven, CT, 06520}

\begin{abstract}

We study the extent of the conformal window for an
$\text{SU}(3)$ gauge theory with $N_f$ Dirac fermions in the fundamental
representation. We present lattice evidence for $12 \leq N_f \leq 16$ that the infrared behavior
is governed by a fixed point, while confinement and chiral symmetry breaking are
present for $N_f \leq 8$.

\end{abstract}


\maketitle


With a small number of massless fermions, a vector-like gauge field theory
such as QCD exhibits confinement and dynamical chiral symmetry breaking. But
if the number of massless fermions, $N_f$, is larger, near but just below the
value, $N_f^\text{af}$, at which asymptotic freedom sets in, the theory is
conformal in the infrared, governed by a weak infrared fixed point (IRFP)
which appears already in the two-loop beta function
\cite{Caswell:1974gg,Banks:1981nn}.  There is no confinement, and chiral
symmetry is unbroken. It is thought that this IRFP persists down to some
critical value $N_f^\text{c}$, where the coupling is sufficiently strong that
the transition to the confined, chirally broken phase takes place. The range
$N_f^\text{af} > N_f > N_f^\text{c}$ is the ``conformal window'', where the
theory is in the ``non-Abelian Coulomb phase''.

Theories in or near the conformal window could play a key role in physics
beyond the standard model.  For example, a theory near the conformal window could describe electroweak symmetry breaking.  It is therefore important to study the extent of this
window, as well as the order of the transition at $N_f^\text{c}$ and the
properties of the theory inside the window and near it. Despite interest in
these questions for many years, little is known with confidence.  This can be
contrasted with supersymmetric QCD, where duality arguments determine the
extent of the conformal window and lead to weakly coupled effective low-energy
theories at both ends \cite{Seiberg:1994pq}.

An upper limit on $N_f^\text{c}$ for both supersymmetric and
non-supersymmetric theories has been proposed based on the counting of
massless degrees of freedom, employing the thermodynamic free energy
\cite{Appelquist:1999hr}.  For a supersymmetric $\text{SU}(N)$ gauge theory
with $N_f$ massless Dirac fermions in the fundamental representation (where
$N_f^\text{af} = 3N$), one finds $N_f^\text{c} \leq (3/2) N$, a limit
precisely saturated by the result from duality arguments.  For a
non-supersymmetric $\text{SU}(N)$ theory with $N_f$ massless Dirac fermions in
the fundamental representation (where $N_f^\text{af} = (11/2)N$), one finds
$N_f^\text{c} \leq 4N [ 1 - (1/ 18N^{2})+ ...]$. It is not known to what
extent this limit is saturated.  The most recent lattice studies
\cite{Iwasaki:2003de}, for $N = 3$, lead the authors to the conclusion that
$N_f^\text{c}$ is much lower.  They find $6 < N_f^\text{c} < 7$.


In this letter, we describe a new lattice study of the conformal window for an
$\text{SU}(3)$ gauge theory with $N_f$ Dirac fermions in the fundamental
representation. We adopt a gauge-invariant definition of the running coupling,
valid for any strength, derived from the Schr\"odinger functional (SF) of the
gauge theory \cite{Luscher:1992an,Sint:1993un,Bode:2001jv}. For an
asymptotically free theory, this coupling agrees with the perturbative running
coupling at short enough distances \cite{Bode:1999sm}.

Making use of staggered fermions as in Ref.~\cite{Heller:1997pn}, it is most
straightforward to restrict attention to values of $N_f$ that are multiples of
$4$. The value $N_f = 16$ leads to an IRFP that is sufficiently weak that it
is best studied in perturbation theory.  The value $N_f = 4$ is expected to be
well outside the conformal window, leading to confinement and chiral symmetry
breaking as with $N_f = 2$.  Thus we focus on $N_f = 8$ and
$N_f = 12$.

The Schr\"odinger functional is the transition amplitude from a prescribed
state at time $t = 0$ to another state at time $t = T$. It can be written as a
Euclidean path integral in a spatial box of size $L$ with Dirichlet boundary
conditions at $t = 0$ and $t = T$ where $T$ is $O(L)$.  Periodic boundary
conditions are imposed in the spatial direction.  The Schr\"odinger functional
can be written as
\beq
 {{\cal{Z}}[W,\zeta, \overline{\zeta}; W^\prime,\zeta^\prime,
  \overline{\zeta}^\prime] =}
   \int [DU  D\chi D\overline{\chi}]
  e^{-S_G - S_F}
\eeq
where $U$ are the gauge fields and $\chi$, $\overline{\chi}$ are the staggered
fermion fields.  $W$ and $W^\prime$ are the (fixed) boundary values of the
gauge fields, and $\zeta, \overline{\zeta}, \zeta^\prime,
\overline{\zeta}^\prime$ are the boundary values of the fermion fields at
$t=0$ and $t=T$, taken here to be zero.  The quantity $S_G$ is the Wilson
gauge action and $S_F$ is the massless staggered fermion
action.

The gauge boundary values $W(\eta), W^\prime(\eta)$ are chosen such that the
minimum action configuration is a constant chromo-electric field \cite{tHooft:1979uj,Luscher:1982ma} whose magnitude is of $O(1/L)$ and is controlled by a dimensionless parameter $\eta$
\cite{Luscher:1993gh}.  The SF running coupling $\overline{g}^{2}(L,T)$ is defined by taking
\beq
  \label{eq:dSdeta}
  \frac{k}{\overline{g}^{2}(L,T)} = \left.
    - \frac{\partial}{\partial \eta} \log \cal{Z}
  \right|_{\eta = 0}\ ,
\eeq

where $k = 12 \left( \frac{L}{a} \right)^2 \left[
   \sin\left( 2\pi a^2/3LT\right)
    + \sin\left( \pi a^2/3LT\right) \right]$ is chosen so that $\overline{g}^{2}(L,T)$ equals the bare coupling at tree level. In general, $\overline{g}^{2}(L,T)$ measures the
response of the system to small changes in the background chromo-electric field.

%
%

For staggered fermions, $L/a$ must be even but $T/a$ must be odd, where $a$ is
the lattice spacing. To cancel the resultant $O(a)$ bulk lattice artifact, the
coupling is defined as the average over $T = L \pm a$:
\beq
  \label{eq:bulk_averaging}
  \frac{1}{\overline{g}^{2}(L)} = \frac{1}{2} \left[
    \frac{1}{\overline{g}^{2}(L,L-a)} + \frac{1}{\overline{g}^{2}(L,L+a)}
  \right].
\eeq
$O(a)$ terms on the Dirichlet boundaries remain \cite{Heller:1997pn}.  We
include a perturbative one-loop counterterm of $O(g_0^4 a)$ in our
calculations to remove partially the $O(a)$ boundary artifact.  Since
${\overline{g}^{2}(L)}$ depends on only one IR scale, $L$, it provides a
technical advantage for studying an IRFP over other possible non-perturbative
definitions of the running coupling, such as from the static potential $V(r)$,
which must be computed from Wilson loops at scales $r \ll L$ to avoid
finite-size effects \cite{Necco:2001xg}.


To set the stage, we review briefly the behavior of ${\overline{g}^{2}(L)}$ in
continuum perturbation theory through three loops. By computing
${\overline{g}^{2}(L)}$ in lattice perturbation theory and setting to zero
terms that vanish as $a/L \rightarrow 0$, a continuum beta function,
$\overline{\beta}$, may be defined such that $L (\partial / \partial L)
\overline{g}^{2}(L) = \overline{\beta}\left(\overline{g}^{2}(L)\right) = b_{1} \overline{g}^{4}(L)
  + b_{2} \overline{g}^{6}(L) + b_{3} \overline{g}^{8}(L) + \cdots$, where the first two (scheme-independent) coefficients are
\beq
  \textstyle
  b_1\!=\!\frac{2}{(4\pi)^2} \left[  11 - \frac{2}{3}  N_f \right] ,
  b_2\!=\!\frac{2}{(4\pi)^4} \left[ 102 - \frac{38}{3} N_f \right] .
\eeq
The third coefficient is scheme dependent, given in the SF scheme by
\cite{Bode:1999sm}
\beq
  b_3^\text{SF} = b_3^{\overline{\text{MS}}} + \frac {b_{2}c_2}{2\pi}
  - \frac{b_{1}(c_3 - c_2^{2})} {8 \pi^2}\ ,
\eeq
where $c_2 = 1.256 + 0.040 N_f$ and $c_3 = c_2^2 + 1.197(10) + 0.140(6) N_f
- 0.0330(2)N_f^2$, and where $b_3^{\overline{\text{MS}}}$ is the three-loop
coefficient defined in the $\overline{\text{MS}}$ scheme (only in the loop
expansion), given by
\beq
\textstyle
  b_3^{\overline{\text{MS}}} = \frac{1}{(4\pi)^6}\left[ \frac{2857}{2}
  - \frac{5033}{18}N_f + \frac{325}{54}N_f^2 \right]\ .
\eeq

For $N_f = 16$, a weak two-loop IRFP exists at $\overline{g}_*^2 \simeq 0.52$.
The higher order corrections are very small in both the SF and
$\overline{\text{MS}}$ schemes.

For $N_f = 12$, the two- and three-loop beta functions also exhibit an IRFP,
although the reliability of the loop expansion is less clear.  The two-loop
IRFP is at $\overline{g}_*^2 \simeq 9.48$. At three loops the fixed point
strength is reduced by roughly $50\%$, to $\overline{g}_*^2 \simeq 5.18\
(5.47)$ in the SF ($\overline{\text{MS}}$) scheme.  (In the
$\overline{\text{MS}}$ scheme, where a four-loop result exists, the fixed
point value increases slightly to $5.91$.)  Since the corresponding loop
expansion parameter $\overline{g}_*^2/4\pi^2$ is smaller than unity,
perturbation theory could provide a reasonable basis of comparison for our
lattice simulations.

For $N_f = 8$, there is no two-loop IRFP. While an IRFP can appear at three
loops and beyond, its scheme dependence and typically large value means that
there is no evidence for an IRFP accessible in perturbation theory. A
non-perturbative study is essential.


With this as background, we next describe our lattice simulations for $N_f =
8$ and $12$.  For $N_f = 8$ we compute $\overline{g}^2(L)$ for bare lattice
couplings $\beta \equiv 6/g_0^2 \in [4.5,7.1]$ with typically $0.1$ spacing.
For $N_f = 12$ we choose $\beta \in [4.15,6.5]$ also with typically $0.1$
spacing \footnote{In these ranges of bare coupling, we do not observe
a bulk phase transition as in Brown et al. \cite{Brown:1992fz}}. In both
cases, we use lattice extents $L/a$ = 4, 6, 8, 10, 12, 16 and 20, with the
larger $L/a$ computations done at fewer $\beta$ values due to the much higher
computational cost.  We use the standard hybrid molecular dynamics (HMD) $R$
algorithm \cite{Gottlieb:1987mq} with unit length trajectories.  We generate
three independent ensembles varying the number of steps per trajectory,
typically in the range 64--128 steps but up to 512 steps at stronger couplings, and perform a quadratic extrapolation to remove finite step size errors.  Within our statistics, the observed systematic shift in $\overline{g}^2(L)$ due to finite step size is negligible over the chosen range of step sizes.

 For the majority of the simulations, those at relatively weak lattice coupling,
we employ of order 40,000 HMD trajectories, sufficient to estimate reliably
autocorrelations. At stronger couplings, the autocorrelations become much
longer, with the time histories showing the previously observed phenomenon
\cite{Luscher:1993gh,DellaMorte:2004bc} of large excursions lasting a few thousand
trajectories.  In such cases, simulations are run longer, up to 80,000 HMD
trajectories. Error estimation is performed using the jacknife method with the block
size adjusted to eliminate the  effects of autocorrelations.

To observe the running of the coupling over a large range of scales requires
the generation of several hundred independent ensembles at various values of
the box size, $L/a$, and bare gauge coupling, $\beta \equiv 6/g_0^2$.  With so
many independent statistical estimates of $\overline{g}^2(L)$, occasional
large statistical fluctuations of these estimates are expected.  So, we model
our estimates with a smooth interpolating function based on a truncated
Laurent series
\beq
  \label{eq:interpolating_function}
  \overline{g}^2(\beta,a/L) = \sum_{k=1}^3
  \frac{c_k(a/L)}{\left[\beta-\beta_0(a/L)\right]^k}\ ,
\eeq
with polynomial dependence of the coefficients on $a/L$
\beq
 \textstyle
  \label{eq:interpolating_function_coefficients}
  c_k = \sum_{l=0}^2 c_{kl} \left( \frac{a}{L} \right)^l\ ,
  \ \beta_0 = \sum_{l=0}^2 \beta_{0l} \left( \frac{a}{L} \right)^l\ .
\eeq
Best-fit values for the coefficients at 8 and 12 flavors will be included in a future paper.
This interpolating function is used only to
describe the lattice data in the limited range where the data exists, well
away from the poles.

The extrapolation to the continuum is implemented using the step-scaling
procedure \cite{Luscher:1991wu,Caracciolo:1994ed}, a systematic method that
captures the renormalization group evolution of the coupling in the continuum limit.  The basic
idea is to match lattice calculations at different values of $a/L$, by tuning
the lattice coupling $\beta \equiv 6/g_0^2$ so that the coupling strength
$\overline{g}^2(L)$ is equal on each lattice. Keeping $\overline{g}^2(L)$
fixed while changing $a/L$ allows one to extract the lattice artifacts.
Previous work by the ALPHA collaboration has shown that perturbative
counterterms greatly reduce $O(a/L)$ artifacts from the running coupling. So,
our preferred method of extrapolating to the continuum limit at each step
assumes that $O(a^2/L^2)$ errors dominate.

In practice, one calculates a discretized version of the running coupling,
known as the step-scaling function. In the continuum, it is designated
\begin{equation}
\sigma(s, \overline{g}^2(L)) = \overline{g}^2(sL),
\end{equation}
where $s$ is the step size.  On the lattice, $a/L$ terms are also present; one
defines $\Sigma(s, \overline{g}^2(L), a/L)$ similarly, so that it reduces to
$\sigma(s, \overline{g}^2(L))$ in the continuum limit:
\begin{equation}
  \sigma(s, \overline{g}^2(L)) = \lim_{a\rightarrow 0}
  \Sigma(s, \overline{g}^2(L), a/L)
\end{equation}
First, a value $u = \overline{g}^2(L)$ is chosen.  Several ensembles with
different values of $a/L$ are then generated, with $\beta \equiv 6/g_0^2$
tuned using our interpolating function Eq.~(\ref{eq:interpolating_function})
so that the measured value of $\overline{g}^2(L) = u$ on each.  Then for each
$\beta$, a second ensemble is generated with larger spatial extent, $L
\rightarrow sL$.  The measured value of $\overline{g}^2(sL)$ on the larger
lattice is exactly $\Sigma(s, u, a/L)$.  Extrapolation in $a/L$ to the
continuum then gives us the value of $\sigma(s,u)$.  Using the value of
$\sigma(s,u)$ as the new starting value, this process may then be repeated
indefinitely, until we have a series of continuum running couplings with their
scale ranging from $L$ to $s^n L$.  In this paper, we take $s = 2$.



\begin{figure}
\includegraphics[width=85mm]{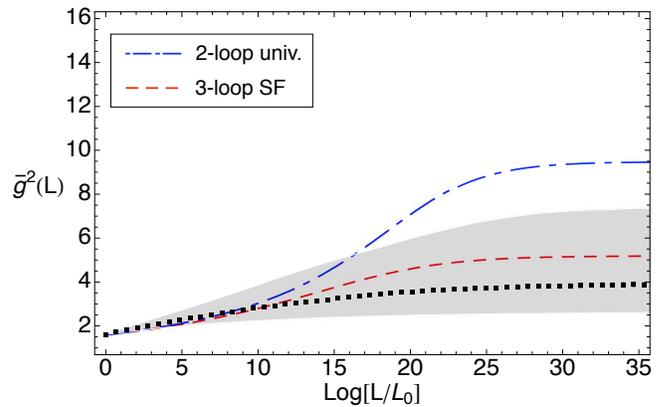}
\caption{\label{fig:nf12_ssf}
Continuum running coupling from step scaling for $N_f = 12$. The statistical
error on each point is smaller than the size of the symbol.  Systematic error
is shown in the shaded band.
}
\end{figure}

Our results for $N_f = 12$ continuum running are presented in
Fig.~\ref{fig:nf12_ssf}. We take $L_0$ to be the scale at which
$\overline{g}^{2}(L) \simeq 1.6$, a relatively weak coupling. The points shown
are for values of $L/L_0$ increasing by factors of two. The step-scaling
procedure leading to these points involves stepping $L/a$ from $4 \to 8$, $6
\to 12$, $8 \to 16$, and $10 \to 20$, and then extrapolating $\Sigma(2,u,a/L)$
to the continuum limit assuming that $O(a^2/L^2)$ terms dominate.

Various sources of systematic error must be accounted for. The interpolating function Eq.~(\ref{eq:interpolating_function}) may not contain enough terms to capture the true form of $\overline{g}^2(L)$ at large $L$ where there is sparse data, and although the $O(a/L)$ terms are expected to be small, ignoring them completely in the continuum extrapolation may introduce a small
systematic effect.  In addition, a few simulations that had run at least 20\% of their target length, but were not yet completed, were included in the fit.  The statistical error in $\overline{g}^2(L)$ for these cases was likely underestimated.  Here we provide an estimate of our systematic error by varying our continuum extrapolation method between extremes.  Inspection of $\Sigma(2,u,a/L)$ as a function of $a/L$ indicates that dropping the step from $4 \to 8$ and performing a constant extrapolation underestimates the true continuum running, while performing a linear fit to all four steps gives an overestimate.  These define the upper and lower bounds of the shaded region in Fig.~\ref{fig:nf12_ssf}, which we take to be a conservative estimate of the overall systematic error.  

The observed IRFP for $N_f = 12$ agrees within the estimated systematic-error
band with three-loop perturbation theory in the SF scheme.  An important
feature for $N_f = 12$ is that the interpolating curves are anchored by values
of $\overline{g}^{2}(L)$ that are also above the IRFP.  For $\beta \leq 4.4$,
$\overline{g}^{2}(L)$ is large, decreasing as $L/a$ increases with fixed $\beta$.  In the step-scaling function, values of $u$ in this range lead to $\Sigma(2, u, a/L) < u$ as $a/L \to 0$.  This behavior
is similar to that found in Ref.~\cite{Heller:1997vh} for $N_f = 16$, and
consistent with approaching the IRFP from above in the continuum limit.  In a
future paper, we will exhibit both the step-scaling results and the continuum evolution in this region.

\begin{figure}
\includegraphics[width=85mm]{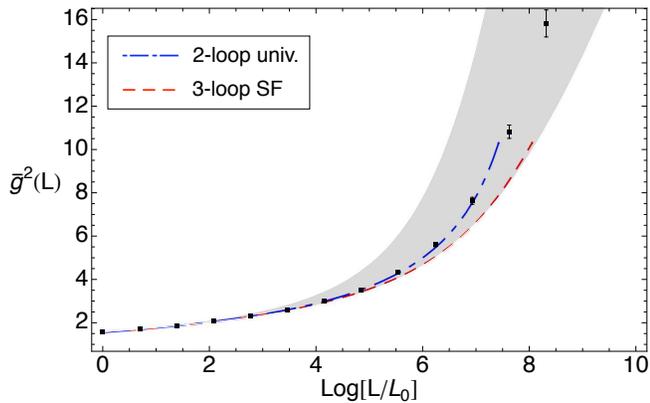}
\caption{\label{fig:nf8_ssf} Continuum running coupling from step scaling
for $N_f = 8$. Errors are shown as in Fig.~\ref{fig:nf12_ssf}. Perturbation
theory is shown up to only $\overline{g}^2(L) \sim 10$. }
\end{figure}

Our results for $N_f = 8$ continuum running are presented in
Fig.~\ref{fig:nf8_ssf}, starting at a scale $L_0$ where $\overline{g}^2(L)
\simeq 1.6$, and exhibiting points with statistical error bars for values of
$L/L_0$ increasing by factors of two. The three step-scaling procedures are
the same as in the $N_f = 12$ case. Stepping $L/a$ from $4 \to 8$, $6 \to 12$,
$8 \to 16$ and $10 \to 20$ with quadratic extrapolation again provides the
points with statistical error bars, and the other two procedures define the
upper and lower bounds of the systematic-error band. For comparison, we have
also shown the results of two- and three-loop perturbation theory up to
$\overline{g}^2 \simeq 10$, beyond which there is no reason to trust the
perturbative expansion.

The $N_f = 8$ running coupling shows no evidence for an IRFP, or even an
inflection point, up through values exceeding $14$. The points with
statistical errors begin to increase above three-loop perturbation theory well
before this value. This behavior is similar to that found for the quenched
theory \cite{Heitger:2001hs} and for $N_f = 2$ \cite{DellaMorte:2004bc},
although, as expected, the rate of increase is slower than in either of these
cases. The coupling strength reached for $N_f = 8$ exceeds rough estimates of
the strength required to trigger dynamical chiral symmetry breaking
\cite{Appelquist:1998rb, Gies:2005as, Gardi:1998ch}, and
therefore also confinement. This conclusion must be confirmed by simulations
of physical quantities such as the quark-antiquark potential and the chiral
condensate at zero temperature.

To conclude, we have provided evidence from lattice simulations that for an
$\text{SU}(3)$ gauge theory with $N_f$ Dirac fermions in the fundamental
representation, the value $N_f$=$8$ lies outside the conformal window, and
therefore leads to confinement and chiral symmetry breaking; while $N_f$=$12$
lies within the conformal window, governed by an IRFP.  We stress that these
conclusions do not depend crucially on the $L/a = 4$ data, which are of limited
use in the SF scheme \cite{Luscher:1992an}.  Thus the lower end of
the conformal window, $N_f^\text{c}$, lies in the range $ 8 < N_f^\text{c} <
12$.

This conclusion, in disagreement with Ref.~\cite{Iwasaki:2003de}, is reached
employing the Schr\"odinger functional (SF) running coupling,
$\overline{g}^2(L)$. This coupling is defined {\it at} the box boundary $L$
with a set of special boundary conditions.  It runs in accordance with
perturbation theory at short enough distances, and is a gauge-invariant
quantity that can be used to search for conformal behavior, either
perturbative or non-perturbative, in the large $L$ limit.

For $N_f$=$8$, we have simulated $\overline{g}^2(L)$ up to values that exceed
rough estimates of the coupling strength required to trigger dynamical chiral
symmetry breaking \cite{Appelquist:1998rb, Appelquist:1997dc, Gies:2005as,
Gardi:1998ch}, with no evidence for an IRFP. For $N_f$=$12$, our observed IRFP
is rather weak, agreeing within the estimated errors with three-loop
perturbation theory in the SF scheme.

The simulations of $\overline{g}^2(L)$ at $N_f$=$8$ and $12$ should be
continued to achieve more precision.  It is also important to supplement the
study of $\overline{g}^2(L)$ by examining physical quantities such as the
static potential, and demonstrating directly that chiral symmetry is
spontaneously broken for $N_f$=$8$ through a zero-temperature lattice
simulation.  Simulations of $\overline{g}^2(L)$ for other values of $N_f$, in
particular $N_f$=$10$, will be crucial to determine more accurately the lower
end of the conformal window and to study the phase transition as a function of
$N_f$. All of these analyses should be extended to other gauge groups and
other representation assignments for the fermions \cite{Sannino:2004qp}.


We acknowledge helpful discussions with Urs Heller, Walter Goldberger, Rich
Brower, Chris Sachrajda and Aneesh Manohar.  This work was supported partially 
by DOE grant DE-FG02-92ER-40704 (T.A.\ and E.N.), and took place partly at the Aspen
Center for Physics (TA).  Simulations were based in part on the MILC code \cite{MILC},
and were performed at Fermilab and Jefferson Lab on clusters provided by the DOE's
U.S.~Lattice QCD (USQCD) program, the Yale Life Sciences Computing Center
supported under NIH grant RR 19895, the Yale High Performance Computing
Center, and on a SiCortex SC648 on loan to Yale from SiCortex, Inc.


\bibliography{QCDConfWind}

\end{document}